\documentstyle[12pt]{article}

\begin{document}

\author{Guang-jiong Ni \thanks{%
E-mail: gjni@fudan.ac.cn} \\
Department of Physics, Fudan University \and Shanghai, 200433, P. R. China}
\title{To Enjoy the Morning Flower in the Evening------Is Special Relativity a
Classical Theory?\thanks{%
This is an English version of its original article published in the Chinese
journal {\it Kexue (Science)} Vol. 50, No. 1, 29$\sim $33,
1998}}
\date{March 10, 1998}
\maketitle

\begin{abstract}
The relation between the special relativity and quantum mechanics is
discussed. Based on the postulate that space-time inversion is equavalent to
particle-antiparticle transformation, the essence of special relativity is
explored and the relativistic modification on Stationary Schr\"{o}dinger
Equation is derived.
\end{abstract}

The 20th century is coming to an end. Many people wish to speculate on the
development of science in the next century. The achievement of physics in
the 20th century is so brilliant that it conceals some mysteries which
remain still since the beginning of the 20th century. Just like the flowers
blossomed in the morning will be more beautiful to enjoy in the evening, we
write down the title, which reminds us of the book ''To pick up the morning
flower in the evening'' by the famous Chinese writer, Lu Xun(1881-1936). I
think we need to learn from Lu in doing the scientific research.

\section{A lesson}

The Special Relativity (SR) was established in 1905. Since then, in more
than 50 years, teachers and students in learning physics were delighted to
talk about the Lorentz contraction------ a rod moving in high speed seems
shorter than it is at rest. It was not until 1959 that J.Terrell pointed out
that the Lorentz contraction is not quite the case that it seems to
be (Phsy. Rev. 116(1959) 1041, see Ref [1] ). The reason is simple. In
general case, the light emitted at the same time from different points of a
moving body can not arrive at the pupil of an observer at a same time. On
the other side, the light which arrive at the pupil at the same time are not
emitted from the body at a same time.

There is a lesson we should learn from the interesting thing mentioned
above. Maybe due to the great success of SR and Quantum Mechanics (QM) , in
the research of theoretical physics as well as in the physics teaching, the
deduction method ( from general case to special case ) is prevailing over
the analysis and induction method ( from special case to general case ). It
is an unbalanced state which does not conform to the law of development of
physics. Surely, the mathematics and logical reasoning are very important.
But one should fuse the deduction and induction methods together, all
depending on the realistic situation under consideration.

\section{Do the SR and QM have the common \newline
essence?}

In various universities, SR is taught either in the course of classical
mechanics or in electrodynamics. It is dealt with as a classical theory,
which is different from the QM in essence. What is the reason? The first
reason is obvious. The QM was established in 1925, 20 years later than the
birth of SR. The second reason is as follows. The phenomena ( as the
mechanical motion of a particle ) discussed in SR all have clear-cut
causality, which seems far from the uncertainty relation or wave particle
dualism in QM.

Perhaps one can raise more reasons. But I do not believe in them at all. The
Maxwell equations were established even earlier than SR for 30 years. But
they are relativistic in essence and need no modification by the
establishment of QM. They remain accurate and effective till now. On the
contrary, the classical mechanics must be reformed radically into
relativistic mechanics before the former can be accommodated in the
framework of SR. Moreover, the classical mechanics can not find its right
place in QM.

\hspace{0.05cm}By contrast, the combination of SR with QM leads naturally to
Relativistic QM ( RQM ) and Relativistic Quantum Field Theory ( RQFT\ ),
which in turn develop into the theory of particle physics, withstanding the
test of numerous experiments.

\hspace{0.03in}\hspace{0.05cm}\hspace{0in}It is said that the reason why the
combination of SR with QM is so productive is due to the complementarity of
classical point of view with quantum point of view. I doubted it quite
early. I thought that, e.g., the marriage of man and woman gives birth to a
child. Are they different humanbeing in essence? No, they must be the same
in essence but different in some aspects. I knew ambiguously that the mating
of different species of livingbeings can not breed descendants.. An exotic
example is the mating of horse with donkey. They breed a mule, but the later
fails to be reproductive again.

Based on the above simple contrast, I firmly believed in the consistency in
essence of SR with QM. After I knew little about that the inheritance gene
is DNA, my problem became the following question:''What are the DNA gained
by RQM\ and RQFT from SR and QM each with 50\%?''

\section{Enlightenment from the discovery of de \newline
'Bloglie's relation}

The de 'Bloglie's relation of a particle , say, an electron, is written in
every textbook on QM. But very few books had explained how de 'Bloglie found
it in 1924. When I found the relevant materials in '70s, I was strongly
attracted and then wrote them into a book [ 2 ].

\hspace{0.14in}\hspace{0.05cm} Surprisingly, the starting point of de
'Broglie was SR. At that time, the quantum concept of light with relevant
Einstein relation

\hspace{1.4cm} \hspace{1cm} 
\begin{equation}
E=hv=\hbar \omega
\end{equation}
\hspace{0in}was well accepted. So as a generalization, de 'Broglie supposed
that Eq.(1) is also valid for an electron. However, a serious contradiction
arises at once. The frequency would increase with the velocity of electron,
a prediction just in opposition to that of slowing down of a moving clock in
SR.

Hence de 'Broglie introduced his second postulate that the frequency
associated with electron is not that of a clock fixed on the electron.
Rather, it is the frequency of a wave accompanying with the electron. The
frequency of wave is measured at a fixed point in space.

\hspace{1pt}When dealing with the wave, the discussion of space-time
transformation in SR needs an alternative invariance as the substitution of
the invariance of the speed of light. Quite naturally, de 'Broglie
introduced the postulate of ``phase invariance'', that is, the phase of
internal clock at any instant is equal to that of wave. de 'Broglie himself
was so emphasizing on this postulate that he called it as the ``law of phase
hamony'' and even considered it as the most basic contribution in his whole
life.( see G. Lochak in ``The Wave-Particle Dualism'' eds by S. Diner et.
al., D. Reidel Publishing Company, 1984 ).

\hspace{3pt}As a next step, de 'Broglie assumed that the group velocity of
wave should equals to the velocity of particle. Thus he derived the relation
uniquely as

$\hspace{3cm}$%
\begin{equation}
\lambda =h/p,\hspace{1cm}p=\hbar k\hspace{1cm}(\hspace{3pt}k=2\pi /\lambda 
\hspace{3pt})
\end{equation}
by means of the Lorentz transformation and the relation between $E$ and $p$:

\hspace{3cm} $\hspace{2cm}$%
\begin{equation}
E^2=m_0^2c^4+p^2c^2
\end{equation}

Therefore, in some sense, de 'Broglie began from the whole theory of SR
combining with the half of quantum theory, Eq.(1), to derive the other half
of quantum theory Eq.(2).

\hspace{0.05cm}After learning his idea, I thought that the consistency in
essence of SR with QM is beyond any doubt. The problem was where is the
breach?

\section{Three points of understanding in learning physics}

\hspace{0.02in}For further discussion, I would like to summarize my three
points of understanding after learning physics for over 40 years.

(a) The observableness of a physical quantity or transformation must be
related, directly or indirectly, to some symmetry ( conservation law ) or
dynamical law. Once such kind of relation is lost, it will cease to be an
observable, i.e., it will lose its meaning.

(b) While there are theorems in mathematics, there are laws as well as
theorems in physics. A quantity in the theorem must be defined unambiguously
and independently in advance. But it is not the case for law, sometimes (
not always ) a law is capable of accommodating a definition of physical
quantity which is not well defined before the establishment of the law.

(c) One of the precious experiences of physics in the 20th century lies in
the fact that a new idea is valuable only if it can find some accurate
mathematical scheme as its carrier. An idea if it can only be expressed by
ordinary language would be probably incorrect, at least it is not deep.

\section{Enlightenment from the discovery of parity violation}

In 1956, the historical discovery of parity violation by T.D. Lee, C.N. Yang
and C.S. Wu et al. initiated a great surge in physics community. I just
graduated from university at that time and could not understand the deep
implication of their work for many years. Succeeding the discovery of
nonconservation of P (parity) and C (particle-antiparticle conjugation
transformation), the nonconservation of CP was found in 1964 whereas the CPT
theorem remains valid. What does it mean ?

According to the point (a) mentioned above, I thought it should be also true
for the ``inversion''. Hence, the nonconservation of C, P, and T
individually implies that the definitions of them are all in trouble.

\section{The time reversal in QM is doubtful}

Quite early, I was doubtful of the so-called time reversal (T inversion).It
is said as follows. First, one uses a camera to record a (microscopic)
process and then show the film in reversed order. If the law of phenomenon
looks quite the same as that shown in normal order, then it is called as the
invariance in time reversal.

Actually, in QM one first changes $t$ into $-t$, then performs
a complex conjugate transformation to get

$\quad $%
\begin{equation}
i
\rlap{\protect\rule[1.1ex]{.325em}{.1ex}}h%
\frac \partial {\partial t}\Psi ^{*}(\overrightarrow{x},-t)=\hat{H}\Psi ^{*}(%
\overrightarrow{x},-t)
\end{equation}

Therefore, the so-called time reversal invariance implies the following
equivalence as stressed in Ref [3]:

\qquad 
\begin{equation}
\Psi (\vec{x},t) \sim \Psi ^{*}(\vec{x},-t)
\end{equation}
which is the precise definition expression by the mathematical language. We
can hardly depend on the camera to see the microscopic particle, let alone
to see it when the time is reversed.

Usually,one finds that the equivalence (5) does hold in the eigenfunction of
stationary state in QM. Now it is found that this equivalence is violated in
the decay of neutral Kaon to the extent of 0.3\%. To our knowledge, no well
accepted explanation is found. But the following possibility is not excluded
that the so-called CP violation may be stemming from a phase angle $\delta $
in the standard model of particle physics. The appearance of $\delta $ is
due to the existence of two kinds of eigenstate ------ state under strong
interaction and that under weak interaction.

It seems to me that if the above explanation could be verified in further
experiments, then CP violation, i.e., nonconservation of T reversal does
display itself as a very special problem in particle physics but has nothing
to do with the basic symmetry in space-time.

Anyhow, after the establishment of SR,we are all agreed that space $%
\overrightarrow{x}$ and time $t$ are treated as equals and mutually
commutative. While the space inversion is defined simply as $\overrightarrow{%
x}\rightarrow -\overrightarrow{x}$ ,why we need an extra complex conjugate
transformation in the definition of time reversal?

\section{What is implied in the nonconservation of C inversion.}

It is interesting to see that there is also a complex conjugation in the
definition of C inversion ( charge conjugation transformation ).If in
accordance with the definition,one performs CPT combined transformation on a
fermion obeying the Dirac equation, then one finds that two complex
conjugation transformations are cancelled. Essentially , only one thing is
done, that is ($\overrightarrow{x}\rightarrow -\overrightarrow{x}$ ,$%
t\rightarrow -t$), [4].

Actually similar situation occurred in RQFT when the CPT theorem was proved
by means of principle of SR. At that time the invariance under ($%
\overrightarrow{x}\rightarrow -\overrightarrow{x},t\rightarrow -t$) was
called as ``invariance of strong inversion''.

The definition of C sounds as if it could be directly perceived. After C
inversion, the electric charge of electron, {\it e}, is changed into -%
{ \it e}, then an electron is transformed into a positron. The concept
behind the definition is that electric charge is something like a fluid, a
positron (electron) carries positive (negative) charge.

The definition of C was quite clear-cut till 1956. Then the experiments
revealed the nonconservation of C inversion in weak interaction processes up
to 100\%. This reflects that the definition of C ceases to be effective, it
does not reflect the real physics. Actually, the concept that ``charge is
something like a fluid '' is no longer valid though it has been existing for
some hundred years.

In the beginning of the 20th century,the charge to mass ratio ($\frac em$)
of an electron was looked as a constant in physics. However, the development
of particle physics reveals that being a measure of coupling strength of
electromagnetic interaction, the fine structure constant $\frac{e^2}{%
\rlap{\protect\rule[1.1ex]{.325em}{.1ex}}h%
c}$ equals to $\frac 1{137}$ at low energy ($\sim m_ec^2=0.51MeV$ ) and
increases to $\frac 1{128}$ at high energy ($\sim m_wc^2 \simeq 80GeV$%
). Since $c$ and $
\rlap{\protect\rule[1.1ex]{.325em}{.1ex}}h%
$ are unchanged universal constants, $e$ is no longer a constant now.
Correspondingly, the ``law of conservation of electric charge'' (which had
made historical contribution when Maxwell introduced the ``displacement
current'' ) is no longer true, its position in physics is already
substituted by the ``law of conservation of electric charge number {\it Q}%
'' ($Q=\frac q{\left| e\right| }$) The latter also corresponds to the ``law
of conservation of (electron) lepton number {\it L}''

An electron has $Q=-1,$ $L=+1$ whereas a positron has $Q=+1,$ $L=-1$. So a
correct ``particle-antiparticle transformation'' has to reflect a
transformation between quantum number 1 and -1.

Surprisingly enough,there was a hint in classical theory long ago. An
electron is moving under the Lorentz force of external electric and magnetic
fields. When it is replaced by a positron, people deemed that the direction
of force is reversed due to the opposite charge $e\rightarrow -e$. However,
there exists an alternative equivalent explanation. It is the mass rather
than the charge which change its sign. We will see below that it is indeed
the case in quantum theory.

\section{Modify the CPT theorem into a postulate}

The sensitive readers surely can guess what we will say below. The
transformation between particle and antiparticle can not be defined
subjectively. It should be a natural consequence of theory under the $%
\overrightarrow{x}\rightarrow -\overrightarrow{x},t\rightarrow -t$
transformation.

Note that in the above space-time inversion, beside the sign change of both $%
\overrightarrow{x}$ and $t$, no complex conjugation is needed, while the
electron turns to positron naturally. Of course, this is a new but very
simple postulate. Look at the plane wave function of an electron:

\begin{equation}
\Psi _{e^{-}}\sim \exp \left\{ \frac i{
\rlap{\protect\rule[1.1ex]{.325em}{.1ex}}h%
}\left( \overrightarrow{p}\cdot \overrightarrow{x}-E\cdot t\right) \right\}
\end{equation}
\hspace{0pt} \qquad \qquad $\left( E>0\right) $, a transformation $%
\overrightarrow{x}\rightarrow -\overrightarrow{x},t\rightarrow -t$ will
bring it to the wave function describing a positron:

\begin{equation}
\Psi _{e^{+}}\sim \exp \left\{ -\frac i{
\rlap{\protect\rule[1.1ex]{.325em}{.1ex}}h%
}\left( \overrightarrow{p}\cdot \overrightarrow{x}-E\cdot t\right) \right\}
\end{equation}
with $\overrightarrow{p}$ and $E(>0)$ being the momentum and energy of
positron.

To our knowledge, Eqs (6) and (7) were first written down by Schwinger et
al. [5]. We had stressed in Ref [6] that the above postulate is of great
importance for the whole theory of RQFT. We see that the charge number 
{\it Q} or lepton number {\it L} is now reflected in Eqs (6) and (7)
and no longer defined by the ordinary language..

What we say implies that we look at the CPT theorem as a postulate, i.e.,
look the theory upside-down. In CPT theorem, one first defined C,P and T,
then proved the invariance of theory with SR as its premise. Now we claim
that the definition of space-time inversion should be simplified to $%
\overrightarrow{x}\rightarrow -\overrightarrow{x},t\rightarrow -t,$ then the
transformation between particle and antiparticle follows naturally, no extra
definition is needed.

Certainly, the correctness of this postulate must be tested by further
experiments. As pointed out at the end of Ref [6], one of the crucial tests
is the derivation of SR from it.

\section{The essence of SR is also the essence of mass generation}

Since the establishment of SR on 1905, some authors have been thinking that
if one could reduce the two ``relativistic principles'' introduced by
Einstein, i.e., the ``principle of relativity''(A) and ``principle of the
constancy of the speed of light''(B) into one principle? Many of them
thought that A must be necessary in SR, it is self evident. So they tried to
discard B by some more ``rigorous'' argument. The historical facts reveal
that such kind of attempt failed again and again. Unfortunately, they had
underestimated the SR too much and they even didn't understand how Einstein
derived the mass-energy relation $\quad E=mc^2$ in 1905.

On the other hand, some theoretical physicists still think that we know
little about the origin of mass. It seems that we need to wait for future
experiments with much higher energy. Mean while, they don't think there is
any problem about the existence of essence in SR.

In our point of view, the mass-energy relation or equation $\,\quad
E^2=m_0^2c^4+p^2c^2$ \quad had taught us a lot of things. Note that the
latter is the relation of a right triangle which implies that the mechanism
for generating the rest mass $m_0$ is ``perpendicular''to that for
generating the ``moving mass'' ( $p/c$ ).

On the other hand, a composite particle comprises many constituent particles
( molecules or atoms ), each with total mass turning into the rest mass of
composite particle. This is a fact which implies that there is a common
essence of mass generation at the level deeper than that of two kinds of
mechanism perpendicular each other. It must be the essence of SR at the same
time.

Mass is some kind of phenomenon at low energy. Let us pick a piece of any
thing at hand. It possesses a rest mass and its mass will increase with its
velocity. We should ask a question as follows:``Why it behaves like this?
What change occurs inside it?'' As this kind of phenomenon is quite general
and simple,so the mass should be originated from one ( not two ) law of
quite general, simple and subtle nature, which in turn is already contained
in the known experimental and theoretical knowledge. The experiments with
much higher energy are not so urgently required for this purpose.

For establishment of SR in 1905, this law was reflected in principles A and
B. They are indivisible.. As a metaphor in ancient Chinese story,Einstein
raised ``A'' just like he painted a ``dragon'' on the wall, then he added
``B'' as the ``eyes'' on the ``dragon'' before the latter can fly out of
wall. At that time he must discussed the coordinate transformation between
two inertial systems with relative motion. Now the objective conditions are
improved much better, we are able to and ought to discuss the problem within
a fixed laboratory system. Then the essential law of mass could be
summarized by the statement mentioned above. ``All kinds of matter contain
two degrees of freedom of both particle and antiparticle states, which
possess the symmetry of mutual transformation under the space-time inversion
( $\overrightarrow{x}\longrightarrow -\overrightarrow{x}$ $,\
t\longrightarrow -t$ )''. We denote the two kinds of state by $\theta $ and $%
\chi $, then it reads:

$\hspace{3cm}\hspace{1cm}\hspace{1cm}$%
\begin{equation}
\theta (-\overrightarrow{x},-t)=\chi (\overrightarrow{x},t)
\end{equation}

The subtlety lies not only on the superficial symmetry, but also on the
relationship of ''my existence inside you whereas your existence inside me
''. When using Eq.(8) as a constraint to construct the motion equation for
electron, we obtain the Dirac equation. Then it is shown that there are both 
$\theta $ and $\chi $ inside the wave function with $\left| \theta \right|
>\left| \chi \right| $. Under this condition the phase of both $\theta $ and 
$\chi $ is like that shown in Eq.(6), not showing the symmetry (8)
explicitly. If we perform a space-time inversion to bring Eq.(6) into Eq.(7)
with $\theta \longrightarrow \chi _{c,}$ $\chi \longrightarrow \theta _c$
and $\left| \chi _c\right| >\left| \theta _c\right| $. Now their phases are
all like that shown in (7) which describes a positron.

When I discovered this point in calculation, I could not prevent myself from
crying:``How subtle it is.'' Nature is full of dialectic. There is a saying
in philosophy that ``The property of matter is determined by the principal
contradiction, especially by its principal aspect''. Here it is precisely
the case. Eq.(8)means that $\theta $ and $\chi $ have opposite phase
evolution directions in space-time. As a metaphor the instinct of $\theta $
is ``eastward'' whereas that of $\chi $ is ``westward''. An electron at rest
is composed of $\theta $ up to 100\%. Once it is set into motion, the
ingredient of $\chi $ will be excited coherently. Because $\left| \chi
\right| <\left| \theta \right| $, $\chi $ is slave whereas $\theta $ is the
master. The former is forced to go ``eastward'' even it is not willing to do
so. What it can do is trying to drag back. In this manner, the inertial mass
of electron, $m$, will increase. The ingredient of $\chi $ will build up
with the increase of velocity $v$ until the limit $v\longrightarrow c$, $%
\left| \chi \right| \longrightarrow \left| \theta \right| $ and $%
m\longrightarrow \infty $.

Meanwhile, the clock associated with the electron is running at a slower and
slower rate. On the contrary, inside a positron $\left| \chi _c\right|
>\left| \theta _c\right| $, the former becomes the master whereas the later
is reduced into the slave. Hence both of them go ``westward''.

The world is really full of contradictions. Loosely speaking, we ourselves
are all not pure. We are matter and antimatter as well, but the latter is
hiding implicitly. The strange effect of SR is nothing but a reflection that
the antimatter which is in a subordinate status is just displaying its
existence tenaciously. With this picture, we reversed the derivation of de
'Broglie and derived the half of SR (kinematics) from other half of SR
(dynamics) in combination with the whole quantum theory [7].

\section{Relativistic modification on the stationary Schr\"{o}dinger equation
}

To me, the above idea with simple calculation was caught more than 20 years
ago. It was impossible to publish them mainly due to my lack of knowledge,
also due to the lack of new result. After doing ``normal'' research for over
20 years, I returned back to this problem and wrote down Ref[7]. Being one
step further, Ref[8] was published. The symmetry (8) was used to QM with two
particle or general many body system,( see also [9] ).

In the past, one solved the stationary Schr\"{o}dinger equation

$\hspace{4.6cm}$%
\begin{equation}
H\psi (\overrightarrow{r_1},\overrightarrow{r_2}...\overrightarrow{r_n}%
)=\varepsilon \psi (\overrightarrow{r_1},\overrightarrow{r_2}...%
\overrightarrow{r_n})
\end{equation}

with eigenvalue $\varepsilon $ which is interpreted as the minus of binding
energy $B$:

$\hspace{4.6cm}$%
\begin{equation}
\varepsilon =-B
\end{equation}

i.e., $\varepsilon $ equals directly to the difference between total energy $%
E$ and the rest energy of n particles $Mc^2$ ( $M=\sum m_{i}$), ( $%
B=Mc^2-E$ ).

Eq.(9) is nonrelativistic and seems difficult to be improved in its non
covariant form for a long time. ( On the other hand, though Dirac equation
is relativistic,it is difficult to generalize to $n \geq 2$ case ).

Now we try to combine the Schr\"{o}dinger equation with symmetry (8).
Interesting enough,we easily find Eq.(9) again but with

$\hspace{4.6cm}$%
\begin{equation}
\varepsilon =\frac 1{2Mc^2}(E^2-M^2c^4)
\end{equation}

Thus the binding energy should be solved from $\varepsilon $ as

$\hspace{4.6cm}$%
\begin{equation}
B=Mc^2[1-(1+\frac{2\varepsilon }{Mc^2})^{\frac 12}]
\end{equation}

Evidently, Eq.(10) is the approximation of (12). Since no Lorentz
transformation is involved in this kind of calculation,which does reflect a
relativistic modification, we are more confident in the consistency of SR
with QM in essence.

\end{document}